\documentclass[twocolumn,preprintnumbers,amsmath,amssymb]{revtex4}
\usepackage{graphicx}% Include figure files
\usepackage{dcolumn}% Align table columns on decimal point
\usepackage{bm}% bold math

\begin{document}

%\preprint{APS/123-QED}

\title{  Quantum critical behavior in heavily doped LaFeAsO$_{1-x}$H$_x$ pnictide superconductors analyzed using nuclear magnetic resonance
 }

\author{ R. Sakurai$^{1}$, N. Fujiwara$^1$\footnote {Email:naoki@fujiwara.h.kyoto-u.ac.jp}, N. Kawaguchi$^{1}$, Y. Yamakawa$^2$, H. Kontani$^2$, S. Iimura$^{3, 4}$, S. Matsuishi$^{3, 4}$, and H. Hosono$^{3, 4}$ }

\affiliation{$^1$Graduate School of Human and Environmental
Studies, Kyoto University, Yoshida-Nihonmatsu-cyo, Sakyo-ku, Kyoto
606-8501, Japan}

\affiliation {$^2$Department of Physics, Nagoya University and JST, TRIP, Furo-cho, Nagoya 464-8602, Japan \ \\
$^3$Material and structures laboratory (MSL), Tokyo
Institute of Technology, 4259 Nagatsuda, Midori-ku, Yokohama
226-8503, Japan \ \\
$^4$Frontier Research Center (FRC), Tokyo
Institute of Technology, 4259 Nagatsuda, Midori-ku, Yokohama
226-8503, Japan
}

%\affiliation{$^3$ Frontier Research Center (FRC), Tokyo Institute of
%Technology, 4259 Nagatsuda, Midori-ku, Yokohama 226-8503, Japan }

%\author{}
%\affiliation{$^6$Frontier Research Center, Tokyo Institute of
%Technology, 4259 Nagatsuda, Midori-ku, Yokohama 226-8503}

%\date{September 5 2011}% It is always \today, today,

%\email { naoki@fujiwara.h.kyoto-u.ac.jp}

\begin{abstract}

We studied the quantum critical behavior of the second antiferromagnetic (AF) phase in the heavily electron-doped high-$T_c$ pnictide, LaFeAsO$_{1-x}$H$_x$ by using $^{75}$As and $^{1}$H nuclear-magnetic-resonance (NMR) technique. In the second AF phase, we observed a spatially modulated spin-density-wave-like state up to $x$=0.6 from the NMR spectral lineshape and detected a low-energy excitation gap from the nuclear relaxation time $T_1$ of $^{75}$As.  The excitation gap closes at the AF quantum critical point (QCP) at $x \approx 0.49$. The superconducting (SC) phase in a lower-doping regime contacts the second AF phase only at the AF QCP, and both phases are segregated from each other. The absence of AF critical fluctuations and the enhancement of the in-plane electric anisotropy are key factors for the development of superconductivity.
\end{abstract}

\maketitle

 The prototypical high-transition-temperature ($T_c$) iron-based pnictide, LaFeAsO$_{1-x}$H$_{x}$ can accept high electron doping and has lead to two new discoveries: (i) H doping leads to a superconducting (SC) phase with double domes [1] following a stripe-type antiferromagnetic (AF) phase, and (ii) further H doping leads to a second AF ordering [2] following the SC double domes. The appearance of the second AF ordering is exotic because one expects a Fermi-liquid state upon heavy carrier doping in high-$T_c$ compounds. Thus, the intriguing question is what is the origin of the second AF ordering? The simplest model is the interband nesting between electron pockets because heavy H doping causes the hole pockets to shrink and the electron pockets to expand [1, 3]. This nesting model predicts an incommensurate spin-density-wave (SDW) state with the nesting vector $\textbf{\emph{q}}=$ ($\pi$, $\frac{\pi}{3}$) or ($\frac{\pi}{3}$, $\pi$) [3], similar to the ($\pi$, 0) or (0, $\pi$) nesting vector between electron and hole pockets leading to stripe-type ordering [4]. The nesting between electron pockets is compatible with superconductivity with s$^{++}$-wave symmetry [5-7]: however, it is not compatible with spin-fluctuation mediated superconductivity with s$^{\pm}$-wave symmetry [8, 9] because it involves a sign change of the SC gap between two electron pockets linked by the nesting vector. Another theoretical investigation that supports spin-fluctuation mediated superconductivity suggests that next nearest-neighbor hopping between iron sites plays a key role in the heavily H-doped regime [10]. The origin of the second AF ordering is deeply associated with SC pairing symmetry.

To elucidate the high-$T_c$ mechanism behind the second AF phase, we focus on quantum criticality around the second AF-phase boundary and investigate electronic states from a microscopic view point using nuclear magnetic resonance (NMR) technique for $^1$H, $^{75}$As, and $^{139}$La.
We prepared 44\%, 49\%, 51\% and 60\% H-doped samples for the present experiments. The H concentration was verified using thermal desorption spectroscopy (TDS). The nominal concentrations for these samples were 53\%, 58\%, 63\%, and 70\%, respectively. In a previous report, we used the nominal concentrations [2], but we use the TDS-determined concentrations for the present work.

The NMR spectra were measured at 35.1 MHz to cover signals from $^{1}$H (I=1/2), $^{75}$As (I=3/2) and $^{139}$La (I=7/2) nuclei with a eight-Tesla superconducting magnet. The relaxation time ($T_1$) for $^{75}$As was measured at 48.2 kOe. The signals at 48.2 kOe come from the powder samples with the basal iron planes parallel to the applied field $\emph{\textbf{H}}$. From these spectra and $T_1$, we found that the 44\% H-doped samples are superconducting, whereas the 49\%, 51\%, and 60\% H-doped samples are antiferromagnetic. The phase diagram determined in the present work is summarized in Figs. 5(b), as described below.

\begin{figure}
\includegraphics{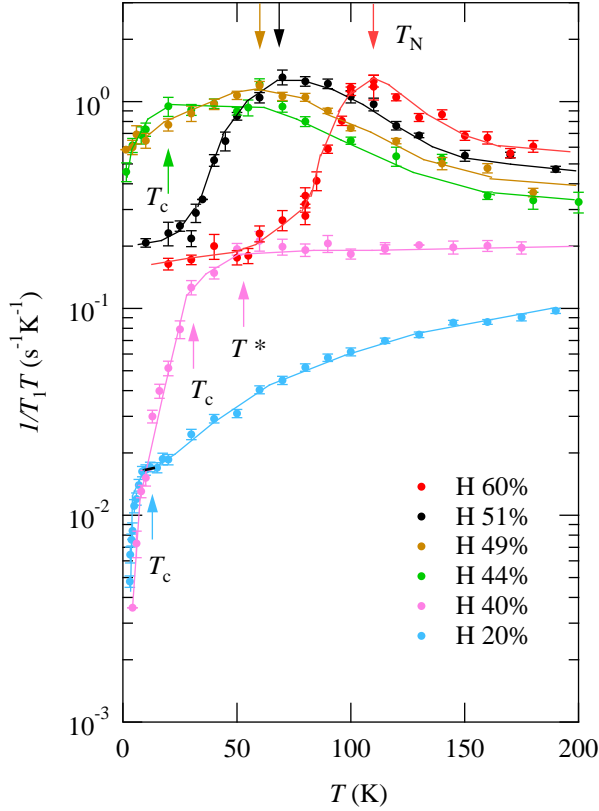}
\caption{\label{fig:epsart} (color online) Nuclear magnetic relaxation divided by temperature $1/T_1T$ for $^{75}$As. $T_c$s indicated by up arrows are determined from the detuning of the NMR tank circuit. $T_N$s are determined from the maximum of $1/T_1T$ as shown by down arrows. The data for 20\% and 40\% H-doped samples are cited from the Ref. [2]. The curves for each doping level are guides to the eye.}
\end{figure}

\begin{figure}
\includegraphics{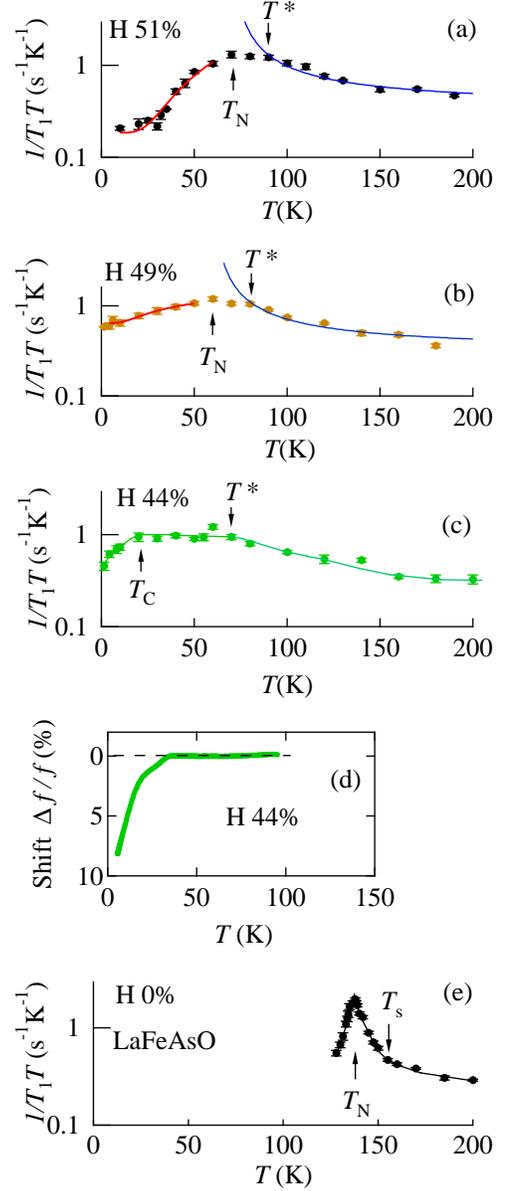}
\caption{\label{fig:epsart} (color online)  (a), (b) $1/T_1T$ in the antiferromagnetic (AF) and paramagnetic (PM) phases.  The data at high temperatures are fitted to the Curie-Weiss law, and the data at low temperatures are fitted to Eq. (1). (c) $1/T_1T$ in the superconducting (SC) and PM phases. (d) Detuning of the NMR tank circuit. It is plotted as the resonance-frequency shift $\Delta f/f$. (e) $1/T_1T$ for the non-doped samples. The data were taken from Ref. [14].  }
\end{figure}

Figure 1 shows the relaxation rate $1/T_1$ divided by temperature ($T$) for various H-doping levels $x$. For the 49\%, 51\% and 60\% H-doped samples, $1/T_1T$ plotted as a function of $T$ forms a broad peak that corresponds to AF ordering. The values of AF-transition temperature $T_N$ are 60, 70, and 110 K, respectively, for the above-mentioned samples. Plots of $1/T_1T$ vs $T$ exhibit systematic variation on H doping, as seen in Figs. 1, 2(a) and 2(b): at high temperatures, $1/T_1T$ for the three H doping levels can be fitted to a Curie-Weiss law [11], reflecting AF fluctuations: $\frac{1}{T_1T}\propto\frac{C}{T-T_N}+a$ where $a$ (= $0.32-0.35$ ($s^{-1}K^{-1}$)) would reflect the density of states. However, for all doping levels, the Curie-Weiss behavior and the AF critical fluctuations near $T_N$ are strongly suppressed with decreasing temperature. In general, the quantity $1/T_1T$ gives a measure of spin fluctuations. $1/T_1T$ involves contributions from all wave vectors $\emph{\textbf{q}}$, $\frac{1}{T_1T} \propto \sum _{q}Im\chi (\emph{\textbf{q}}, \omega)/\omega$, and thus the divergence of $\chi(q, \omega)$ for specific $\textbf{\emph{q}}$ is strongly suppressed. The deviation from the Curie-Weiss behavior suggests the emergency of a pseudo gap. We denote the onset temperature as $T^*$.

Below $T_N$, $1/T_1T$ exhibits activated $T$ dependence:
\begin{equation}
 \frac{1}{T_1T} \propto \alpha + e^{-\Delta/T}
\end{equation} where the activation energy $\Delta$ decreases from 710 to 40 K with H-doping level decreasing from 60\% to 49\%.  The first term represents a gapless excitation predominant at low temperatures and is proportional to the square of the residual density of states. The fitting curves are shown in Figs.2(a) and 2(b). The equation (1) is reminiscent of a SDW state; In a SDW state, $1/T_1T$ is expressed by using nth logarithmic function Li$_n$($x$) as $1/T_1T$=c$_0$+c$_1$$\Phi$($x$) where $\Phi$($x$)=$1/x^2$Li$_1$($e^{-x}$)+$1/x^3$Li$_2$($e^{-x}$) and $x=\Delta/T$ [12,13]. The $T$ dependence of $1/T_1T$ in the heavily H-doped samples clearly contrasts that in the lightly F-doped samples [14-16]: the critical fluctuations accompanied by a sharp upturn toward $T_N$ appears, and moreover, structural transition temperature $T_s$ can be identified from $T$ dependence of $1/T_1T$ (See Fig. 2(e)). The appearance of the critical fluctuations suggests that  $\chi(q, \omega)$ makes a sharp peak at $\emph{\textbf{q}} \approx $ ($\pi$,0) or (0, $\pi$).

Now consider $x=0.44$. Suppression of AF fluctuations, namely pseudo-gap behavior is manifested as a plateau in $1/T_1T$ ranging from 70 to 20 K followed by the SC gap at low temperatures. The onset temperature is shown as $T^*$ in Fig. 2(c). The pseudo-gap behavior is also observed in $1/T_1T$ for $x$=0.40 where explicit Curie-Weiss behavior is absent, as shown by $T^*$ in Fig. 1. The plateau of $1/T_1T$ above $T^*$ would reflect the density of states, and therefore the pseudo-gap behavior, which is widely observed at temperatures above $T_c$ or $T_N$, is attributable to the suppression of AF fluctuations and/or the density of states. For $x=0.44$, superconductivity with $T_c=20 K$ emerges accompanied by a drop in $1/T_1T$. The appearance of superconductivity was also confirmed from the detuning of the NMR tank circuit, which is plotted as the resonance-frequency shift in Fig. 2(d). However, for $x=0.49$, the $T$ dependence of $1/T_1T$ in Fig. 2(b) shows no indication of superconductivity. These results suggest that the AF and SC phases are segregated from each other in the electronic phase diagram. The lack of successive SC and AF transitions upon decreasing temperature excludes the possibility of homogeneous coexistence of AF and SC states, unlike  Ba(Fe$_{1-x}$Co$_x$)$_2$As$_2$ [17, 18] and Ca(Fe$_{1-x}$Co$_x$)AsF [19].

$^{75}$As and $^{139}$La spectra give information of the AF spin configurations. Figure 3(a) compares the first and second AF phases, and Figs. 3(b) and 3(c) compare AF and paramagnetic (PM) phases. The lineshape of $^{75}$As depends on the in-plane anisotropy  of the electric field gradient $\eta  \equiv \frac{V_{yy}-V_{xx}}{V_{zz}}$: the lineshape at 120 K for heavily H-doped samples is asymmetric because of large $\eta$ ($\approx 0.6$), whereas that for lightly F-doped samples has a two-peak structure because of small $\eta$ ($\approx 0-0.1$).

\begin{figure}
\includegraphics{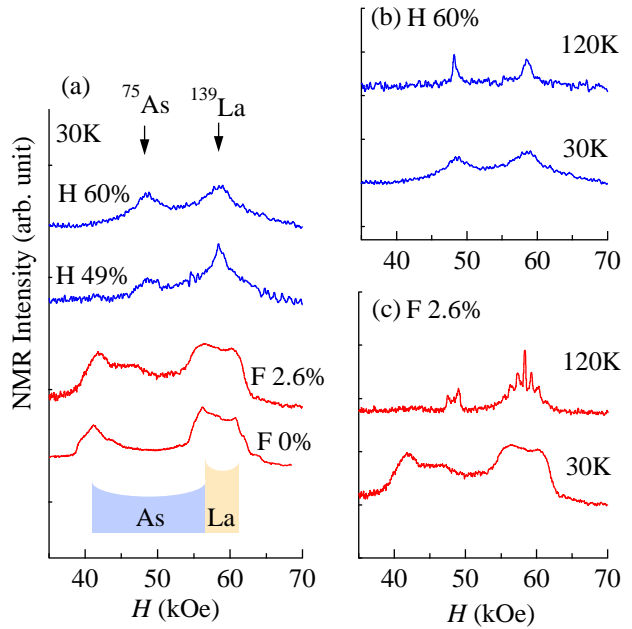}
\caption{\label{fig:epsart} (color online) $^{75}$As and $^{139}$La NMR spectra measured at 35.1MHz. (a)  Spectra in the AF phases for   LaFeAsO$_{1-x}$F$_{x}$ and LaFeAsO$_{1-x}$H$_{x}$. The illustration indicates powder patterns of the AF phase for the lightly F-doped samples. A part of the data for the F-doped samples are taken from Ref. [14]. (b) and (c) Comparison of spectra between the AF and PM phases. }
\end{figure}

\begin{figure}
\includegraphics{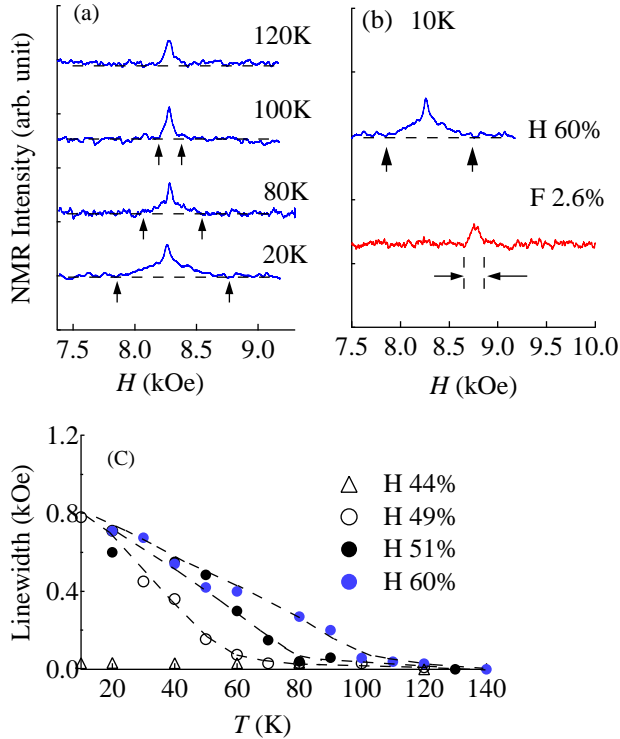}
\caption{\label{fig:wide} (color online)  $^1H$-NMR spectra for 60\% H-doped samples. (a) Spectra at several temperatures below and above $T_N$ ($=100 K$). (b) Comparison of $^1$H-NMR spectra in LaFeAsO$_{1-x}$H$_x$ with $^{19}$F-NMR spectra in LaFeAsO$_{1-x}$F$_x$. (c) The $^1$H linewidth of LaFeAsO$_{1-x}$H$_x$ for several doping levels. The natural linewidth observed in a PM state was subtracted in the figure.  }
\end{figure}

In the first AF phase, the signals from both $^{75}$As and $^{139}$La are rectangular, as seen in Fig. 3(a), reflecting uniform AF moments. A rectangular powder pattern is derived without any specific conditions [19]. For small amounts of F doping, the spectra were unchanged except for a hump at 45 kOe that comes from minor PM (or SC) domains [14]. Unlike the spectra for the first AF phase, those for the second AF phase were of cusp-type up to $x$=0.6, suggesting that spin moments are spatially modulated or a SDW state is realized throughout the second AF phase. The cusp-type pattern comes from summing the rectangular-type patterns with different $\textbf{\emph{H}}$ spans.

\begin{figure}
\includegraphics{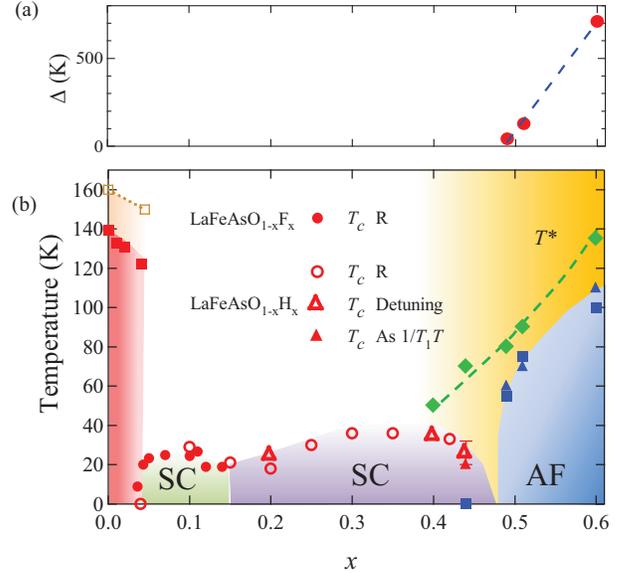}
\caption{\label{fig:epsart}  (color online) (a) Low-energy AF-excitation gap determined from Eq. (1) as a function of H concentration $x$. (b)Phase diagram of LaFeAsO$_{1-x}$F$_x$ and LaFeAsO$_{1-x}$H$_x$. The structural transition temperatures ($T_s$) [14,21] and the antiferromagnetic (AF) transition temperatures ($T_N$) [4, 21] of LaFeAsO$_{1-x}$F$_x$ are shown by open and closed squares, respectively. The superconducting transition temperatures ($T_c$) were determined from resistivity measurements (closed and open red circles) [1, 20], detuning of the NMR tank circuit (open red triangles), and $1/T_1T$ (closed red triangles). $T_N$s of LaFeAsO$_{1-x}$H$_x$ were determined from $1/T_1T$ (closed blue triangles) and the linwidth (closed blue squares). Pseudo-gap behavior is seen at temperature below $T^*$. The yellow-shaded regime shows the enhancement of the in-plane electric-field-gradient anisotropy $\eta$.  }
\end{figure}

One can estimate the value of the maximum spin moments by comparing $^1$H and $^{19}$F spectra. Figure 4(a) shows $^1$H spectra for the 60\% H-doped samples at several temperatures, and Fig. 4(b) compares $^1$H and $^{19}$F spectra. We denote the linewidth as the $\textbf{\emph{H}}$ span between the arrows. The linewidth for $^1$H reaches 1 kOe at 10-20 K. The stripe-type uniform moments ( $\approx$ 0.36 $\mu_B$ [4]) induce an internal field of 0.2 kOe at $^{19}$F sites; hence, the maximum spin moment is estimated to be five times 0.36 $\mu_B$, or 1.80$ \mu_B$. Figure 4(c) shows the difference in the linewidths between AF and PM states, and $T_N$s determined from the linewidth agree well with those obtained from $1/T_1T$. Following the NMR results, the recent neutron-scattering measurements confirmed the existence of the second AF phase and suggested a uniform spin configuration with spin moments of 1.22$\mu_B$ [22]. Regarding high-energy fluctuations, an inelastic-neutron-scattering peak was observed at $\textbf{\emph{q}}$=($\pi$, 0.35$\pi$) for $x$=0.4, which suggests an incommensurate SDW state [23]. The discrepancy of how the spin configuration looks is currently an open problem.

The phase diagram obtained from NMR measurements is summarized in Fig. 5(b). As seen in Fig. 5(a), a fairly large AF-excitation gap defined by Eq. (1) likely closes at $x\approx0.49$; hence, the AF QCP is deduced as $x\approx0.49$, whereas an AF QCP is undefined for the first AF phase [20, 21]. The electronic state at the AF phase boundary gives a clue to determine pairing symmetry. Homogeneous coexistence of AF and SC states around the first AF-SC phase boundary more likely occurs for s$^{\pm}$-wave symmetry than for s$^{++}$-wave symmetry [24$-$27]. Our present work shows no indication of the coexistence of the AF and SC order parameters, and the SC phase contacts the AF phase only at the AF QCP. In this respect, the superconductivity with s$^{++}$-wave symmetry would be more advantageous than that with s$^{\pm}$-wave symmetry, if the above-mentioned theories are applicable for the heavily H-doped regime.

By comparing the two SC-AF phases in Fig. 5(b), one finds that the second SC dome has a higher $T_c$ and a wider doping regime than the first SC dome. For this phenomenon, both spin and orbital properties seem to play key roles. Unlike the case of the first AF phase, AF critical fluctuations are absent in the second AF phase; hence, the superconductivity likely develops in a wider $x$ region that is free from strong AF fluctuations. As for the orbital properties, the SC domes favor large $\eta$ because the second SC dome is close to the large-$\eta$ region as shaded in yellow in Fig. 5(b). The enhancement of $\eta$ suggests the differentiation of d-orbital occupation weight or orbital ordering. The enhancement of $\eta$ and the emergence of the pseudo gap occurs (See $T^*$ in Fig. 5(b)) in the same doping region, which implies that the suppression of strong AF fluctuations and/or the density of states is deeply associated with the orbital degrees of freedom, or the orbital ordering. In this work, we have focused on heavy H doping in LaFeAsO$_{1-x}$H$_{x}$. Recently, another SC double-dome structure was observed in LaFe(As$_{1-x}$P$_x$)(O$_{1-y}$F$_y$) [28$-$30]. A systematic study would increase understanding of high-$T_c$ mechanisms in pnictides.

In summary, we studied the quantum critical behavior of the second AF phase for LaFeAsO$_{1-x}$H$_x$. The AF spin configuration is a spatially modulated SDW-like state up to $x$=0.6. We detected the low-energy AF-excitation gap in the second AF phase: the gap of 710 K at $x$=0.6 closes at the AF QCP ($x\approx0.49$). The AF and SC phases contact each other only at the AF QCP, and both phases are segregated from each other. The suppression of AF fluctuations and the enhancement of in-plane electric anisotropy emerge in the same doping region. These quantities are associated with the pseudo-gap behavior and the orbital degrees of freedom, respectively, and are key factors to explain why the second SC dome manifests itself over a wider doping regime compared with the first SC dome.

\ \\

The NMR work is supported by a Grant-in-Aid (Grant No. KAKENHI 23340101) from the Ministry of Education, Science, and Culture, Japan. This work was supported in part by the JPSJ First Program. We thank T. Morinari, K. M. Kojima and R. Kadono for discussion.

%\ \\

%\noindent {\bf Fig. 1.} \ \\

%\noindent {\bf Fig. 2.} \  \\

%\noindent {\bf Fig. 3.}  \ \\

%\noindent {\bf Fig. 4.} \ \\

%  \ \\

\end{document}